\documentclass{jpsj3}

\title{Novel Magnetic Phases Revealed by Ultra-High Magnetic Field in 
the Frustrated Magnet ZnCr$_{2}$O$_{4}$}

\author{Atsuhiko Miyata}
\inst{Institute for Solid State Physics, University of Tokyo, 5-1-5, Kashiwanoha, 
Kashiwa, Chiba, 277-8581, Japan}
\inst{Department of Applied Physics, University of Tokyo, 7-3-1, Hongo, 
Bunkyo-ku, Tokyo, 113-8656 Japan}
\author{Hiroaki Ueda}
\inst{Institute for Solid State Physics, University of Tokyo, 5-1-5, Kashiwanoha, 
Kashiwa, Chiba, 277-8581, Japan}
\author{Yutaka Ueda}
\inst{Institute for Solid State Physics, University of Tokyo, 5-1-5, Kashiwanoha, 
Kashiwa, Chiba, 277-8581, Japan}
\author{Yukitoshi~Motome}
\inst{Department of Applied Physics, University of Tokyo, 7-3-1, Hongo, 
Bunkyo-ku, Tokyo, 113-8656 Japan}
\author{Nic Shannon}
\inst{H. H. Wills Physics Laboratory, University of Bristol, Tyndall Av., BS8-1TL, 
United Kingdom}
\author{Karlo Penc}
\inst{Research Institute for Solid State Physics and Optics, H-1525 Budapest, 
P.O.B. 49, Hungary}
\author{Atsuhiko Miyata$^{1, 2}$, Hiroaki Ueda$^{1}$\thanks{Present address: Department of Chemistry, Graduate School of Science, Kyoto University, Kyoto 606-8502}, Yutaka Ueda$^{1}$, Yukitoshi Motome$^{2}$, Nic Shannon$^{3}$, Karlo Penc$^{4}$, Shojiro Takeyama$^{1}$\thanks{E-mail address: takeyama@issp.u-tokyo.ac.jp}}
\inst{Institute for Solid State Physics, University of Tokyo, 5-1-5, Kashiwanoha, 
Kashiwa, Chiba, 277-8581, Japan$^{1}$

Department of Applied Physics, University of Tokyo, 7-3-1, Hongo, 
Bunkyo-ku, Tokyo, 113-8656 Japan$^{2}$

H. H. Wills Physics Laboratory, University of Bristol, Tyndall Av., BS8-1TL, 
United Kingdom$^{3}$

Research Institute for Solid State Physics and Optics, H-1525 Budapest, 
P.O.B. 49, Hungary$^{4}$}
\abst{The Faraday rotation technique is used to map out the finite-temperature phase diagram of the prototypical 
frustrated magnet ZnCr$_{2}$O$_{4}$, in magnetic fields of up to 190 T generated by  the single-turn coil method.
We find evidence for a number of magnetic phase transitions, which are well-described by the theory based on spin-lattice coupling.    
In addition to the 1/2 plateau and a 3:1 canted phase, a 2:1:1 canted phase is found for the first time in chromium spinel oxides, which has been predicted by a theory of Penc {\it et al.} to realize in a small spin-lattice coupling limit. Both the new 2:1:1 and the 3:1 phase are regarded as the supersolid phases  according to a magnetic analogy of Matsuda and Tsuneto, and Liu and Fisher. }

\kword{Faraday rotation method,  single-turn coil method, ultra-high magnetic field, geometrically frustrated magnet, magnetic supersolid, spin-lattice coupling}

\begin{document}
\maketitle

\section{Introduction}
The search for new phases of quantum matter is as central to modern condensed 
matter physics as the historical search for new elementary particles.
In recent years, this search has focused increasingly on frustrated magnet systems in 
which conventional descriptions of magnetic order break down entirely.  
%




%
ZnCr$_{2}$O$_{4}$ is a prototypical example of such a frustrated magnet.
Despite strong (\mbox{$J \approx 52 $ K}) antiferromagnetic exchange interactions between 
neighbouring Cr$^{3+}$ ions, ZnCr$_{2}$O$_{4}$ fails to order magnetically down to a 
temperature of \mbox{$T_N = 12.5$ K}, just few percent of the \mbox{$T_{\sf MF} \approx  400$ K} 
anticipated from mean field theory~\cite{lee00,ueda06}.
The unconventional nature of magnetism in ZnCr$_{2}$O$_{4}$ raises the possibility that 
it might also exhibit new forms of order in magnetic field.
Indeed, it has long been known that the interplay between magnetic field and frustrated
exchange interactions can lead to both magnetization plateau --- magnetic ``solids'' 
which break the translational symmetries of the lattice, and spin-flopped phases --- magnetic 
``superfluids'' with well-defined magnetic order in the plane perpendicular to the magnetic field.
One particularly intriguing proposal is that magnetic field might stabilize phases which
break both sets of symmetries simultaneously.   
Such a phase would be a magnetic analogue of the ``supersolid'' long sought 
in $^4$He~\cite{balibar10}.
%


%
For a long time, testing these predictions for materials like ZnCr$_{2}$O$_{4}$ was rendered
impossible by the need for multiple extremes --- magnetic fields $B$ in excess of 100 T and temperatures
of a few Kelvin.
Now, a new generation of pulsed high-field magnets and advances in instrumentation 
make it possible to explore this physics for the first time.  
Here we report the first determination of the magnetic phase diagram of ZnCr$_{2}$O$_{4}$, 
through Faraday rotation, for magnetic fields of up to 190 T, and temperatures down to 4.2 K.
We find evidence for three distinct magnetic phase transitions, the 1/2 plateau and a 3:1 canted phase, a 2:1:1 canted phase, all of which are
well-described by a simple model based on spin-lattice coupling.    
The 2:1:1 canted phase is found for the first time in ZnCr$_{2}$O$_{4}$ as a series of chromium spinel oxides. This phase has been predicted by a theory of Penc {\it et al.} to realize in a small limit of the spin-lattice coupling. The new 2:1:1 phase is regarded as the reentrant supersolid phase after the 3:1 phase appeared in this system,  according to an exact magnetic analogy proposed by  Matsuda and Tstuneto~\cite{matsuda70}, and Liu and Fisher~\cite{liu73}.
%
The finite temperature evolution of these phases is studied, and compared with 
Monte Carlo simulations of an effective spin model.  
These results provide a beautiful illustration of the phenomenon of ``order by distortion''~\cite{tchernyshyov02}, 
the subtle interplay between spin and lattice degrees of freedom in a frustrated magnet.
\section{Theoretical Background}


\begin{figure}
\includegraphics[width=\columnwidth]{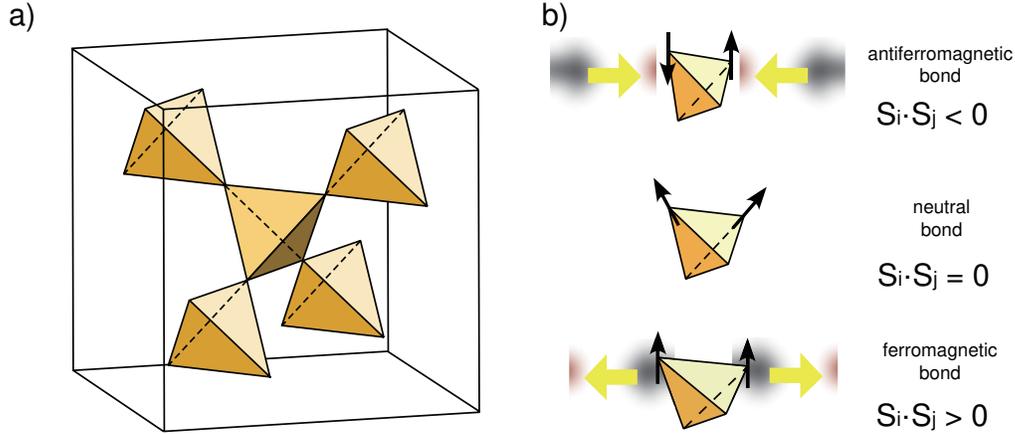}
\caption{\label{fig:lattice} 
(Color online) (a) Pyrochlore lattice of magnetic  Cr$^{3+}$ ions within the spinel oxide ZnCr$_{2}$O$_{4}$, 
shown in the cubic, high-temperature, unit cell.
Cr$^{3+}$ ions with magnetic moment 3 $\mu_B$ are situated at the corners of each tetrahedron.
(b) Simple model of spin-lattice coupling in Cr spinels : antiferromagnetic bonds with 
${\bf S}_i\cdot {\bf S}_j < 0$ contract, while ferromagnetic bonds with 
${\bf S}_i\cdot {\bf S}_j > 0$ extend, in order to maximize the energy gain from exchange interactions.    
This systematic distortion of individual bonds lifts the frustration inherent in the pyrochlore lattice, 
stabilizing unconventional magnetic ground states.
}
\end{figure}


The spinel oxide, ZnCr$_2$O$_4$ 
is an electrical insulator which bears all the hallmarks of a frustrated magnet.
%
The metallic Cr ions are in a $3+$ ionization state, with a magnetic moment of 3 $\mu_B$, and occupy the sites of a 
highly-frustrated pyrochlore lattice, built of corner-sharing tetrahedra (cf. Fig.~\ref{fig:lattice}).  
At high temperatures, the magnetic susceptibility of ZnCr$_2$O$_4$ shows Curie-Weiss law 
behaviour, with a Curie temperature of $\theta_{CW} \approx 390$ K\cite{lee00,ueda06}, suggesting that 
neighbouring Cr$^{3+}$ ions have antiferromagnetic exchange interactions \mbox{$J \approx 52$ K}.
However in the absence of magnetic field, ZnCr$_2$O$_4$ does not order magnetically 
unless cooled to $T_N = 12.5$ K, just a few percent of $\theta_{CW}$~\cite{lee00,ueda06}.   
The onset of this magnetic order is accompanied by a distortion of the lattice from cubic to tetragonal~\cite{lee00}.   


The figure of merit $T_N/\theta_{CW} \approx 0.03$, 
suggests that ZnCr$_2$O$_4$ is one of best known approximations to an ideal (classical) frustrated antiferromagnet with a pyrochlore lattice possessing a three dimensional antiferomagnetic nearest neighbor interaction, for which $T_N = 0$, and hence 
$T_N/\theta_{CW} \equiv 0$~\cite{reimers92,moessner98}.  
It is therefore a very attractive system to study in magnetic field.  
Such high-field experiments have already been performed for the 
sister compounds HgCr$_2$O$_4$~\cite{ueda06} and CdCr$_2$O$_4$~\cite{kojima08}.  
%
These show a dramatic half--magnetization plateau with $m=3/2$ $\mu_B$~\cite{ueda05,ueda06,mitamura07,kojima08}.   
Again, the transition into the half--magnetization plateau is accompanied by a structural distortion 
of the lattice.


It is clear that magnetic order in the $A$Cr$_2$O$_4$ spinels 
is intimately linked to lattice structure.
In fact the main features of the magnetic phase diagrams of HgCr$_2$O$_4$ and CdCr$_2$O$_4$ 
are well-described by a simple theory due to Penc {\it et al.}, 
which describes the spin-lattice coupling within the tetrahedra making up the pyrochlore 
lattice~\cite{penc04,motome06,penc07,shannon10}, see also ref. 16.   
Since exchange interactions originate in the overlap of electronic orbitals, magnetic energy can be gained 
by shortening antiferromagnetic bonds and lengthening ferromagnetic ones (cf. Fig.~\ref{fig:lattice}).  
This effect, known as magnetostriction, plays a secondary role in conventional magnets.
However it is essential to understanding Cr spinels, where it lifts the degeneracy associated with 
the pyrochlore lattice, and has been dubbed ``order by distortion''~\cite{tchernyshyov02}.



In its simplest form, magnetostriction can be accounted for by an effective, classical, spin model 
\begin{eqnarray}
    \label{eqn:H}
     \mathcal{H} &=& J \sum_{\langle ij \rangle} \big[ {\bf S}_i \cdot {\bf S}_j
       - b \, ({\bf S}_i \cdot {\bf S}_j)^2 \big]
	   -  h \sum_i S^z_i \, ,
\end{eqnarray}
where spin-lattice coupling is parameterized by the biquadratic interaction $b$, 
$\langle ij \rangle$ counts the nearest-neighbour bonds of a pyrochlore lattice, 
and $h$ is the external magnetic field, measured in natural units~\cite{tchernyshyov02, penc04}.
%
This effective spin model can easily be generalized to take into account longer range exchange 
interactions, and has the great advantage that it is accessible to Monte Carlo simulation.
The different ordered states found can be classified according to the {\sf A}$_1$, {\sf T}$_2$ and 
{\sf E} irreducible representations of the tetrahedral symmetry group $\cal{T}\it{_d}$~\cite{penc04,penc07}. 
For large values of $b$, eq.~(\ref{eqn:H}) favours collinear states, including
a collinear half-magnetization plateau, in which three spins point up, and one down, 
within each tetrahedron (a state with local {\sf T}$_2$ symmetry)~\cite{penc04}.   


Empirically, the strength of the spin-lattice coupling $b$ is found to depend on the 
metallic $A$-site ion.   
Experiments on HgCr$_2$O$_4$ and CdCr$_2$O$_4$, for 
which spin-lattice coupling is strong, reveal dramatic half-magnetization plateaux
and magnetic phases diagrams in strikingly good agreement with simulations of 
eq.~(\ref{eqn:H}), for values of $b \approx 0.1$~\cite{motome06}.
Experiments under pressure confirm that $b_\text{Cd} < b_\text{Hg}$, and suggest that 
spin-lattice coupling is weaker again in ZnCr$_2$O$_4$~\cite{ueda08,jo05}.


For small values of $b$, eq.~(\ref{eqn:H}) predicts a still-richer phase diagram, dominated
by coplanar phases which simultaneously break spin-rotation symmetry in the 
plane perpendicular to the magnetic field, and exhibit modulation in the 
component of the spin, aligned with it.   
These phases are a natural generalization of the 
magnetic supersolids proposed by Matsuda and Tsuneto~\cite{matsuda70} and Liu and Fisher~\cite{liu73}.  
These and subsequent authors investigated magnetic supersolid stabilized by the combination of magnetic field 
and easy-axis anisotropy. 
However the theory of Penc {\it et al.}~\cite{penc04} suggests that the combination of magnetic field and 
spin-lattice coupling provides a new route to supersolid phases, even in the absence of magnetic 
anisotropy.
In particular, for small values of spin-lattice coupling, it predicts an unusual 2:1:1 canted phase bordering 
on the half-magnetization plateau, which is a magnetic supersolid.
This state
is not present for the larger values of spin-lattice coupling realized in 
HgCr$_2$O$_4$ and CdCr$_2$O$_4$, and has yet to be observed in experiment.
%
Could this phase be found in ZnCr$_2$O$_4$?

\section {Experimental Procedures}


\begin{figure}
\includegraphics[width=\columnwidth]{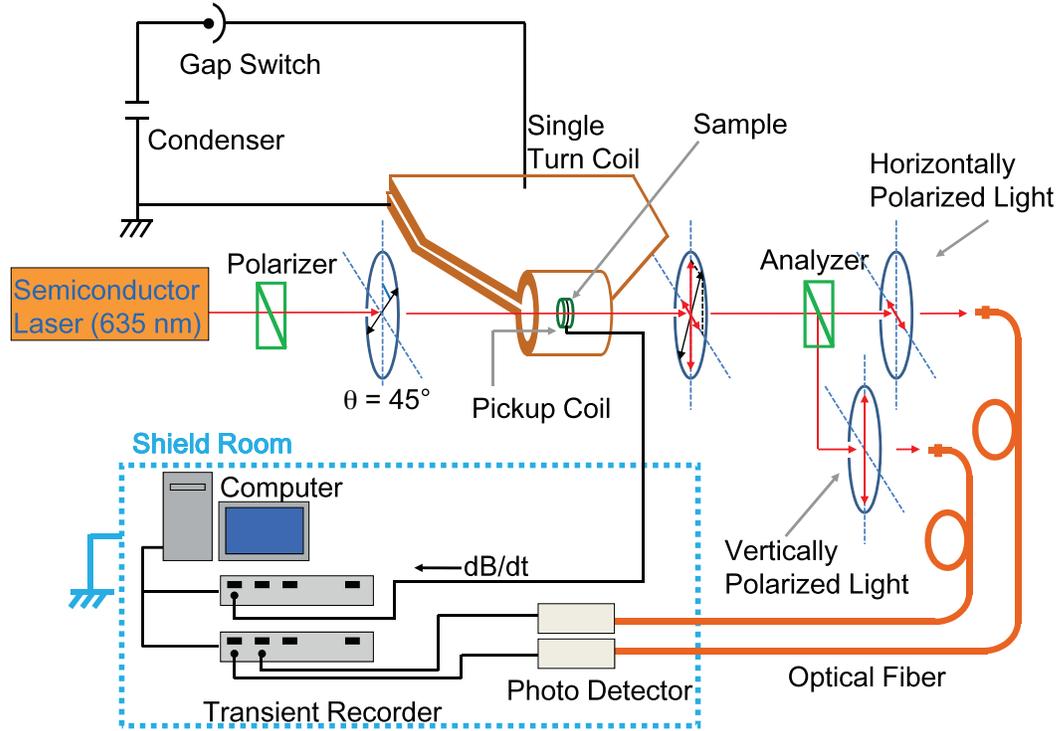}
\caption{\label{fig:setup} 
(Color online) Schematic plan of experiments to measure the magnetization of ZnCr$_2$O$_4$ in high 
magnetic field.
Extremely high magnetic fields are generated by discharging a bank of capacitors through 
a single-turn coil.
The sample is illuminated with light from a 
laser, linearly polarized at 
$45^{o}$ to the vertical axis.  
The polarization of this light undergoes Faraday rotation through an  
angle which depends on the magnetization of the sample.    
Horizontally and vertically polarized components of the transmitted light are resolved, 
and used to reconstruct the magnetization of the sample during the magnetic field pulse.
The sample is held at a constant temperature of between 4.2 and 24 K by a liquid-He flow type
cryostat (not shown).
}
\end{figure}


\begin{figure}
    \includegraphics[width=.7\textwidth]{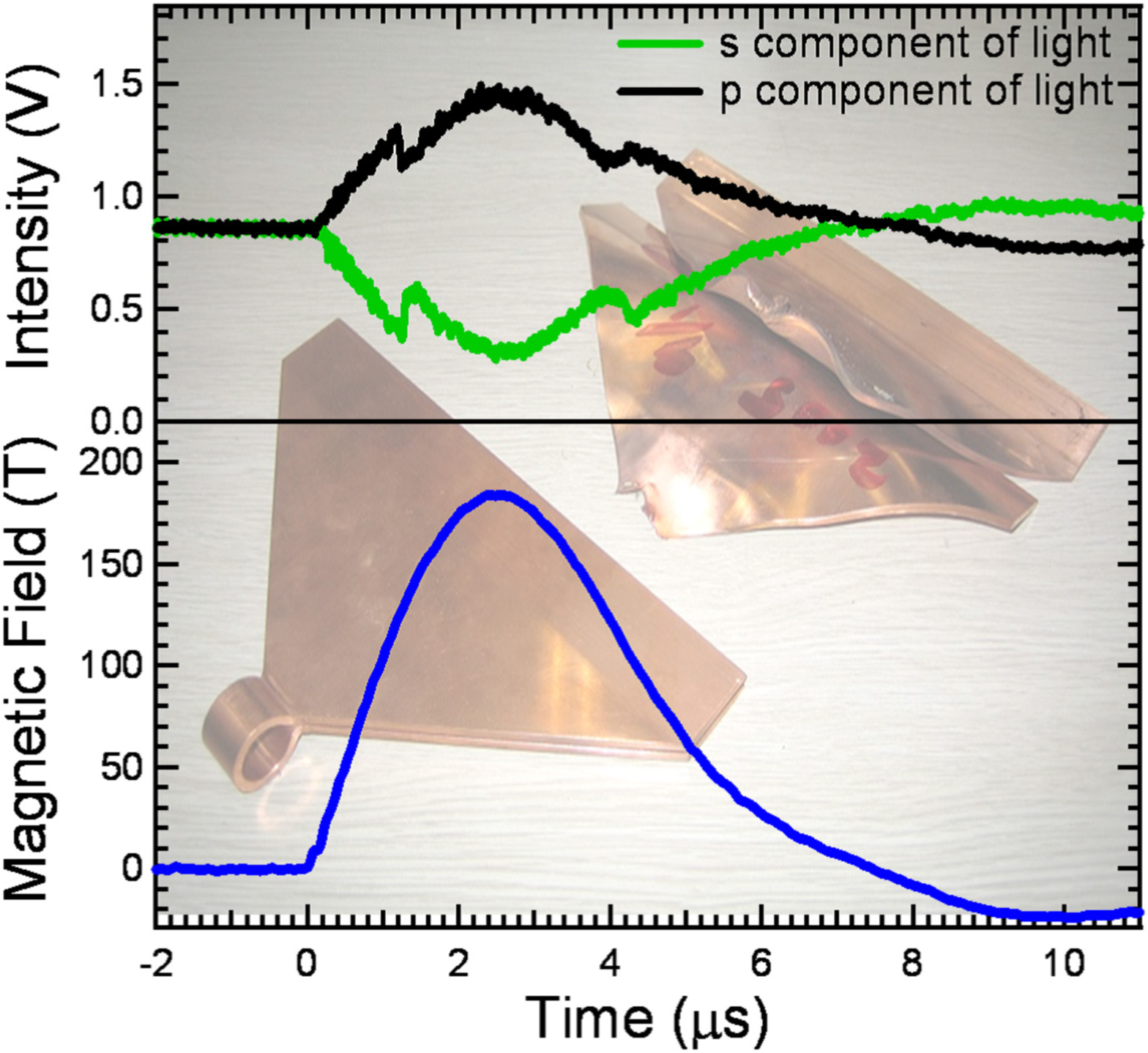}
 \caption{
\label{fig:faraday-rotation}     
(Color online) Profile of magnetic field (lower panel), and intensity of transmitted light 
(upper panel) as a function of time, during a 190~T magnetic field pulse 
generated by a single-turn coil.
Light is resolved into vertical (s) and horizontal (p) components with intensity $I_s$ and $I_p$, and the 
angle of the rotation of the light is then given  as $\theta_F = \frac{1}{2} \sin^{-1}(\frac{I_p-I_s}{I_p+I_s})$. 
Since the Faraday angle $\theta_F$ is directly proportional to the magnetization $m$ of the sample, 
this leads to the magnetization curve shown in Fig.~\ref{fig:mofh}.
%
%
The magnetic field curve was obtained by integrating the voltage induced in a 
separate pick-up coil.
The backdrop to these plots shows the solid copper single-turn coil before 
and after the experiment.     
}
\end{figure}


Measurements of ZnCr$_2$O$_4$ ($T_N/\theta_{CW} \approx 0.03$) in magnetic field are clearly desirable, since it is 
a better approximation to an ideal frustrated pyrochlore antiferromagnet than either HgCr$_2$O$_4$ ($T_N/\theta_{CW} \approx 0.18$)~\cite{ueda06} or CdCr$_2$O$_4$ ($T_N/\theta_{CW} \approx 0.11$)~\cite{ueda05}, and posses a relatively small spin-lattice coupling.  
However the scale of the antiferromagnetic interactions in ZnCr$_2$O$_4$ immediately present 
a problem, since the saturation field for an ideal pyrochlore antiferromagnet with $J \approx 52$ K 
is of order $400$ T --- well beyond the reach of any conventional magnet. 
Such fields are however accessible through electro-magnetic flux compression techniques developed recently
at the Institute for Solid State Physics of the University of Tokyo, which can generate up to $700$ T~\cite{takeyama10}.   
%
%
Moreover, the low value of the N\'eel temperature $T_N=12.5$ K in the absence of field suggests that 
experiments must be performed at temperature of a few Kelvin.  
To achieve these multiple extremes, we have used a hand-made liquid-He flow type cryostat made totally of a ``stycast'' resin~\cite{takeyama87}, 
which fits within the single-turn copper coil of a pulsed magnet, capable of producing magnetic fields 
approaching 200 T for a duration of $\sim 10$ $\mu$s~\cite{nakao85,herlach73}.   
In these experiments, the inner diameter of the single-turn coil was $10$~mm.  
A current of the order of $2$--$3$~MA was injected into this coil by discharging $200$~kJ fast capacitor banks charged to $50$~kV. The value of magnetic field during the pulse was measured by a calibrated pick-up coil located in the vicinity of the sample as shown in Fig.~\ref{fig:setup}. The calibration of a pick-up coil is described in detail in ref. 22. 
The estimated error of the absolute value of the magnetic field found by this method is about $\pm $3 $\%$. 
The lower panel of Fig.~\ref{fig:faraday-rotation} shows the waveform of a $190$~T magnetic field pulse generated by the single-turn coil method. 


The remaining challenge is to determine how the magnetic properties of ZnCr$_2$O$_4$ change with magnetic
field, within the microsecond timeframe of the pulsed field.  
Under such extreme conditions, using a traditional electro-magnetic induction method to measure the magnetization 
of the sample has many associated difficulties. 
The greatest of these arises from the huge background voltage induced in the pick-up coil\cite{mitamura07,takeyama88}.
This problem can be avoided by measuring the magnetization optically, using the Faraday rotation method.
The experimental setup is shown schematically in Fig.~\ref{fig:setup}.  A semiconductor laser (a coherent ``Cube'') of  the wavelength 635 nm was used as 
a light source.   
The light incident on the sample was linearly polarized at $45^{o}$ to the vertical axis.  
The transmitted light was separated into perpendicular, linearly polarized s and p components by a 
Wollaston prism, as shown in Fig.~\ref{fig:setup}.
Raw data for the intensities $I_s$ and $I_p $ of the transmitted light are shown in Fig.~\ref{fig:faraday-rotation}. 
The magnetization of the sample was calculated from these as $M \propto \theta_F = \frac{1}{2} \sin^{-1} \left(\frac{I_p-I_s}{I_p+I_s}\right)$.
Independent measurements in a long-pulse magnet for fields of up to 50 T
demonstrate that the induced magnetization of the sample can be accurately measured by this method, once 
appropriate allowance has been made for the diamagnetism of the sample and quartz substrate \cite{kojima08}.
In this earlier work, we have demonstrated that the Faraday rotation method correctly reproduces independent measurements
of the sample magnetization for fields of up to $50$~T, and used it to measure the complete magnetization curve of CdCr$_2$O$_4$ 
in magnetic fields of up to 140 T~\cite{kojima08}.

A single crystal sample of ZnCr$_2$O$_4$ was grown by a vapour transport technique. 
The sample was cut parallel to the (111) crystal surface, attached on a quartz substrate and polished to about 
100 $\mu $m thickness.
The diameter of the resulting sample was about 1.5 mm. 
The magnetic field was applied parallel to the [111] direction. This direction is equivalent in all existing domains of ZnCr$_2$O$_4$, because there are three domains in which hard axes are along [100], [010], or [001] directions below T$_N$~\cite{kikuchi02, yoshida06}.
%
%


Monte Carlo simulations were performed using a local-update Metropolis algorithm to sample spin configurations.
We typically perform $10^6$ MC samplings for measurements after $10^5$ steps for thermalization.
We have checked the convergence by comparing the results for different initial spin configurations. 
In particular, to minimize the hysteresis associated with first-order transitions, we used mixed initial conditions 
in which different parts of the system are assigned different ordered or disordered states.

\section{Results and Discussion}
\subsection{Identification of novel magnetic supersolid phase}


\begin{figure}
    \centering
    \hfill
    \includegraphics[width=.43\textwidth]{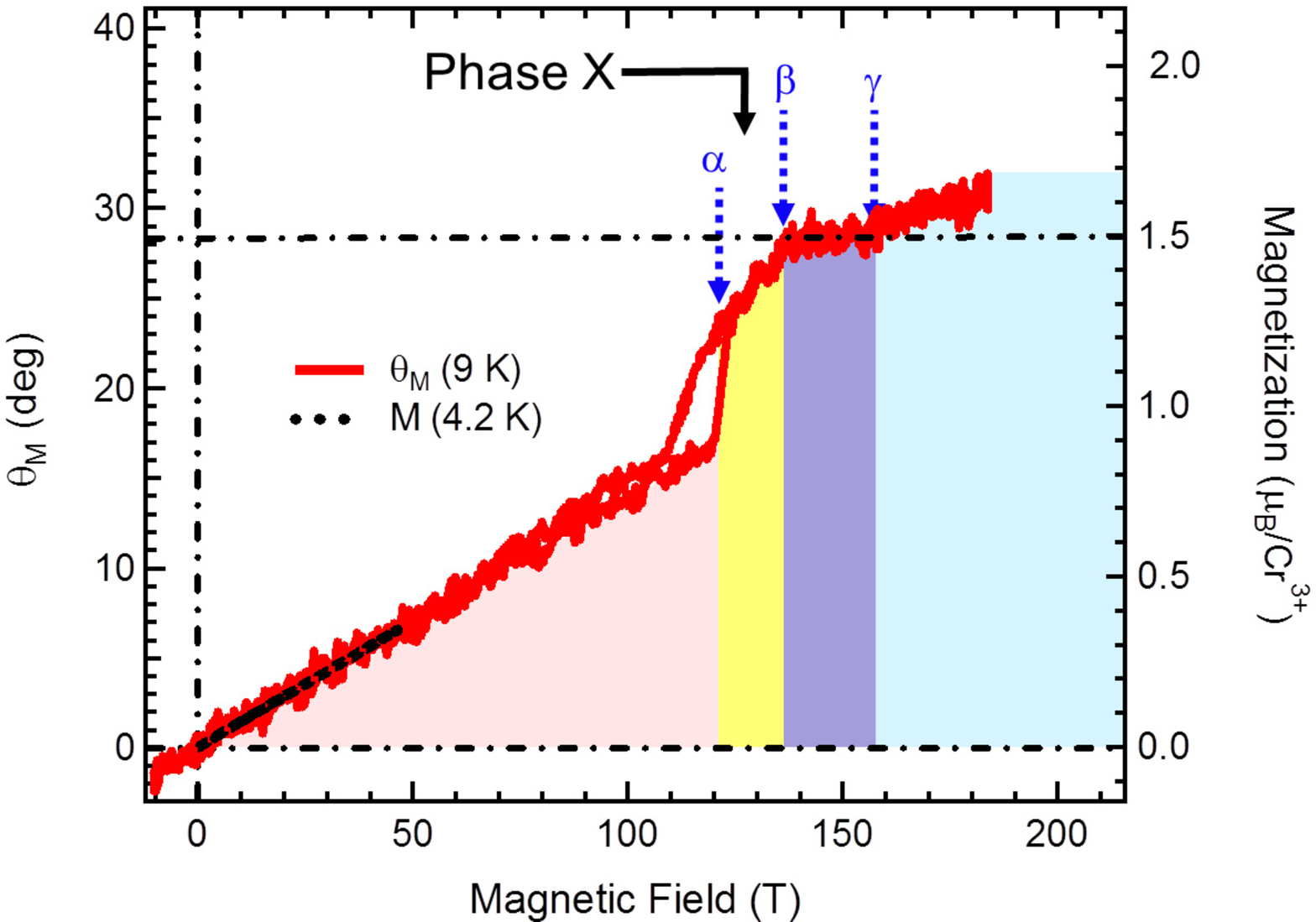}
    \hfill
    \includegraphics[width=.37\textwidth]{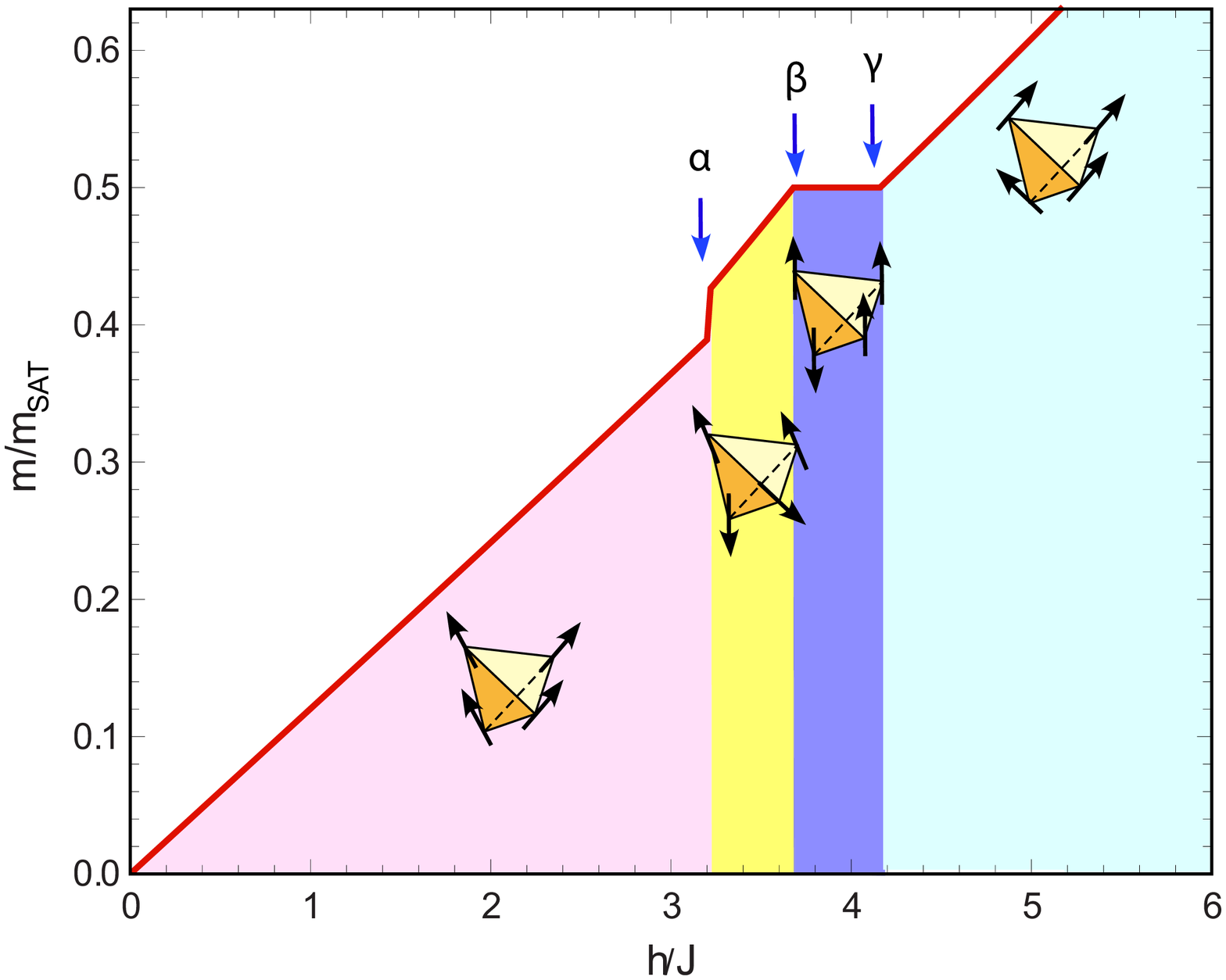}
    \hfill\null
\caption{\label{fig:mofh} 
(Color online) (a) The magnetization of ZnCr$_2$O$_4$ in a magnetic field of up to 190 T, as determined
by Faraday rotation, at a temperature of 9 K (solid red line).
The Faraday rotation $\theta_M$ has been corrected by a constant offset 
$\theta_M = \theta_F - \delta \theta_{\sf dia.}$, to allow for the diamagnetism
of the sample and quartz substrate.
A magnetisation in absolute units [$\mu_B$/Cr$^{3+}$] is obtained by normalizing 
to an independent measurement made in a long-pulse magnet for fields 
of up to 50 T, at a temperature of 4.2 K (black dashed line).
Magnetic phase transitions inferred from this data are labelled
with Greek characters ($\alpha $, $\beta $, and $\gamma $).
(b) Theoretical prediction for the magnetization of a Cr spinel with antiferrimagnetic 
nearest neighbour exchange $J$, and weak spin-lattice coupling $b=0.02$, in applied 
magnetic field, following ref. 12.   
Spin-lattice coupling relieves the frustration associated with the pyrochlore lattice, 
stabilizing new magnetic phases in applied magnetic field.   
The spin configurations  within a single tetrahedron are shown for each phase.
}
\end{figure}


In Fig.~\ref{fig:mofh}(a), we present results for the magnetization of ZnCr$_2$O$_4$ at a temperature of $9$~K, 
for fields $B$ of up to $190$~T.    
The magnetic phase transitions identified from this data are labelled $\alpha$, $\beta$ and $\gamma$.   
For $B < B_\alpha = 120$ T the magnetization is linear in field, as would be expected for an antiferromagnet
with canted spins.    
In the theory of Penc {\it et al.}, shown in Fig.~\ref{fig:mofh}(b) for a spin-lattice coupling $b=0.02$, 
this phase is a 2:2 canted state in which spins are canted in pairs within each tetrahedron ({\sf E} symmetry state).    


At $B_\alpha = 120$ T the system undergoes an abrupt, first-order transition into a state with higher 
magnetization and finite, field-independent magnetic susceptibility, labelled ``Phase~X'' in Fig.~\ref{fig:mofh}(a). 
We identify this phase with the previously unobserved 2:1:1 canted phase predicted by Penc {\it et al.} 
for small values of spin-lattice coupling --- cf. Fig.~\ref{fig:mofh}(b). 
This unusual spin configuration is stabilized by the competition between spin and lattice degree of freedom, 
and mixes the {\sf T}$_2$ and {\sf E} symmetries of a single tetrahedron.
Moreover, since the modulation of the $S^z$ component of the canted spins breaks the space group symmetries 
of the lattice (``solid'' order), while their $S^x$ and $S^y$ components break spin rotation symmetry 
(``superfluid'' order), this phase is a magnetic supersolid.  
This novel supersolid phase persists up to \mbox{$B_\beta = 135$ T}, when the systems undergoes a continuous phase
transition in a state with vanishing magnetic susceptibility and exactly half the saturation magnetization.    
This is the collinear half-magnetization plateau with a local {\sf T}$_2$ symmetry, previously seen in 
CdCr$_2$O$_4$~\cite{ueda05,kojima08} and HgCr$_2$O$_4$~\cite{ueda06}.    
At $B_\gamma = 158$ T the system undergoes a second continuous phase transition into another phase
with finite magnetic susceptibility.   
We identify this with the 3:1 canted {\sf T}$_2$ state predicted by Penc {\it et al.}, which is also a magnetic supersolid.



Within the theory of Penc {\it et al.}, the ratio $\omega = W/B_\gamma$ of the onset field $B_\gamma$ for the half-magnetization plateau
to its width $W$ provides a measure of the effective spin-lattice coupling $b$.
For HgCr$_2$O$_4$, $B_\gamma = 10$ T and $W = 17$ T ($\omega_\text{Hg} = 1.7$), 
while in CdCr$_2$O$_4$, in $B_\gamma = 28$ T and $W = 34$ T ($\omega_\text{Cd} = 1.2$).
In comparison, in ZnCr$_2$O$_4$, $B_\gamma = 135$ T, and $W = 23$ T ($\omega_\text{Zn} = 0.17$).
It follows that $\omega_\text{Hg} > \omega_\text{Cd} \gg \omega_\text{Zn}$.  
This is entirely consistent with observation of a 2:1:1 canted state and the finding from pressure experiments 
that the spin lattice coupling in Cr spinels becomes stronger as the size of the A-site ion is increased, 
i.e. $b_\text{Zn} \ll b_\text{Cd} < b_\text{Hg}$~\cite{ueda08,jo05}.    


\subsection{Evolution of phases at a finite temperature}


\begin{figure}
\includegraphics[width=\columnwidth]{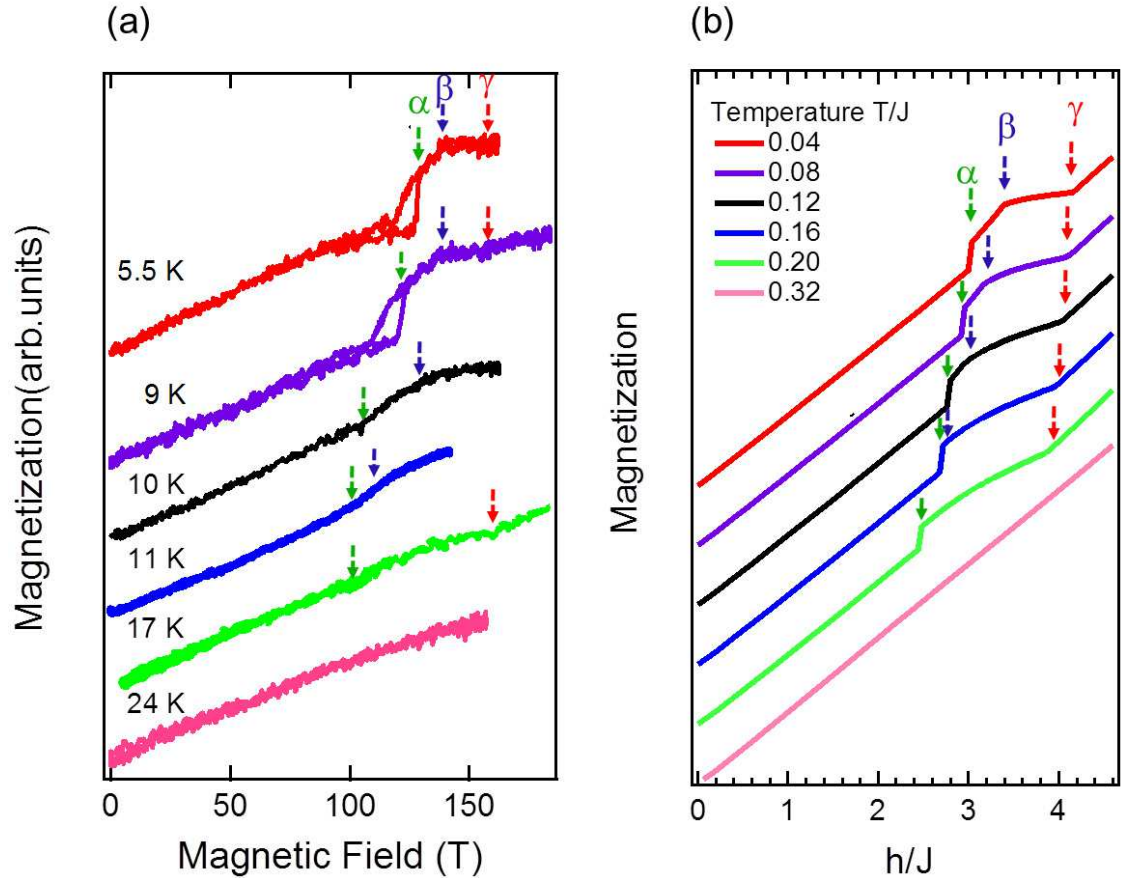}
\caption{\label{fig:finite-T} 
(Color online) (a) Magnetization of ZnCr$_2$O$_4$ obtained from Faraday rotation in fields of up to 190 T, 
for temperatures ranging from 5.5 K to 24 K.   
Arrows show the location of magnetic phase transitions.
Errors in the measurement of temperature and magnetic field are estimated to be 
$\pm 1$ K and $\pm 3$ $\%$, respectively.
(b) Magnetization of bilinear-biquadratic Heisenberg model with $b=0.02$, 
 as a function of magnetic field $h$, as determined by Monte Carlo simulation.
This model captures the essential features of spin-lattice coupling in Cr spinels.
Simulations are performed for temperatures ranging from $0.04 J$ to $0.32 J$, 
where the nearest neighbour exchange $J \approx 52$ K in ZnCr$_2$O$_4$.
Arrows show the location of magnetic phase transitions.
}
\end{figure}


\begin{figure}
\includegraphics[width=\columnwidth]{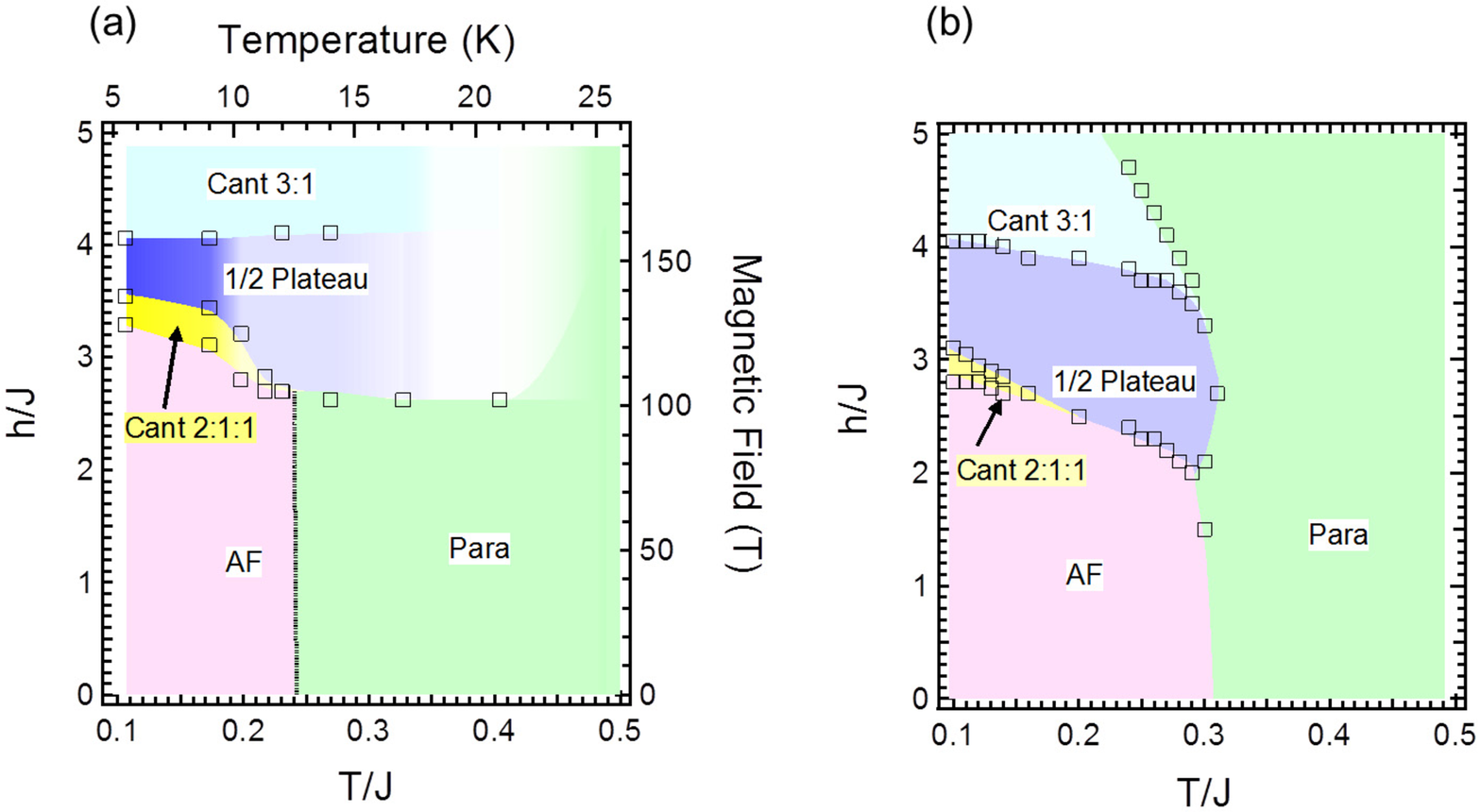}
\caption{\label{fig:phase-diagram} 
(Color online) (a) Experimental magnetic phase diagram of ZnCr$_2$O$_4$ as found by Faraday rotation. 
(b) Theoretical prediction for magnetic phase diagram of ZnCr$_2$O$_4$, as determined by Monte Carlo
simulation of a bilinear-biquadratic Heisenberg model capturing the essential features 
of spin-lattice coupling.
The parameter for this model are the same as those used to calculate the magnetization
in Fig.~\ref{fig:finite-T}(b).
}
\end{figure}


Measurements of the magnetization 
of ZnCr$_2$O$_4$ were made for  a range of temperatures, illustrated in Fig.~\ref{fig:finite-T}(a), 
and showed a strong temperature dependence.   
For comparison, in Fig.~\ref{fig:finite-T}(b), we show the predictions of eq. (\ref{eqn:H}), taken
from Monte Carlo simulations with $b=0.02$.
These experimental results permit to construct a temperature-field phase diagram for 
ZnCr$_2$O$_4$.   
This is shown in Fig.~\ref{fig:phase-diagram}(a), where temperature (magnetic field) has been 
normalized to the value of $J=52$ K  ($J=38.9$ T)\cite{We}.


An equivalent phase diagram can also be constructed from these Monte Carlo simulations, where
phase transitions are identified from peaks in the relevant order parameter susceptibilities.
The results for spin-lattice coupling $b=0.02$ are shown in Fig.~\ref{fig:phase-diagram}(b), 
where (arbitrary) four-sublattice order was enforced by the inclusion of a ferromagnetic 
third-neighbour interaction $J_3 = -0.05$ \cite{while}.
Overall the agreement between theory and experiment is excellent, considering the 
small number of adjustable parameters used ($b$ and $J_3$). 
Critically, the topology of the experimental phase diagram is correctly reproduced
by the theory, with the new phase identified with a 2:1:1 state (a magnetic supersolid) 
sandwiched between the 2:2 canted antiferromagnet and half-magnetization plateau.
The main difference between theory and experiment is seen in the absolute value 
of the transition temperatures $T_N$ from the paramagnetic phase to the antiferromagnet 
or half-magnetization plateau phase.
However these are highly sensitive to longer-range interactions, which in turn will be 
modified by the structural transitions associated with each magnetic phase
transition --- an effect not treated in the theory.


It is also hard to give precise estimations of critical fields from experiment, since small 
changes in the magneto-optical data are hard to resolve at higher temperatures.
However we observed a strong temperature dependence of critical fields, as shown in Fig.~\ref{fig:phase-diagram}(a). 
The width of the half-magnetization plateau state widens as temperature increases, as a result of stabilization 
of collinear states by thermal fluctuations. 
This is an example of the entropy--driven ``order by disorder'' effect~\cite{kawamura85,zhitomirsky02}, and it 
becomes more evident at the limit of small spin-lattice coupling.
%


\section{Conclusions}


We have used Faraday rotation to establish the magnetic phase diagram of ZnCr$_2$O$_4$ 
in pulsed fields of up to $190$ T, for temperatures down to $4.2$ K.   
Four distinct magnetically ordered phases can be distinguished at low temperatures; 
these include a half-magnetization plateau, and three phases with canted spins and a finite magnetic susceptibility.
The key features of these results are all well-described by a simple model based on spin-lattice coupling.
On the basis of this analysis, we find that spin-lattice coupling in ZnCr$_2$O$_4$ stabilizes
two canted phases bordering on the half-magnetization plateau, both of which are magnetic supersolids 
in the sense of Matsuda and Tsutneto or Liu and Fisher. 
One of these phases is an unusual 2:1:1 canted state, predicted to occur in the limit of small spin-lattice 
coupling, but never previously observed.
We conclude that the combination of magnetic field and spin-lattice coupling can stabilize novel magnetic 
supersolid phases in frustrated magnets, even in the absence of easy-axis anisotropy.     

\section*{Acknowledgments}
This work was supported by a Grant-in-Aid for Scientific Research on Priority Area High 
Field Spin Science in 100 T (No. 451), 
a Grant-in-Aid for Scientific Research (No. 19052004), 
a Grant-in-Aid for Scientific Research (No. 20740190)
a Grant-in-Aid for Scientific Research on Priority Areas (No. 19052008) from the Ministry of Education, 
Culture, Sports, Science, and Technology (MEXT) of Japan, 
UK EPSRC grants EP/C539974/1, EP/G031460/1 and EP/G049483/1, 
and Hungarian OTKA Grant No. 73455.



\begin{thebibliography}{99}
\bibitem{We}
We consider $J k_B = J g \mu_B = 4.5$ meV, 
with a Land\'e g-factor of $g=2$.
\bibitem{while}
While the precise form of long range order 
cannot be inferred from magnetization alone, the theory is remarkably robust against different 
long range interactions~\protect\cite{shannon10}.
\end{thebibliography}
\end{document}